\documentstyle[12pt,agums]{article}
\lefthead{Ma and Summers}
\righthead{Formation of Power-law Energy Spectra}
\received{August~6,~1998}
\revised {September~28,~1998}
\accepted{September~30,~1998}
\paperid{}
\cpright{AGU}{1998}
\ccc{0148-0227/98/98GL-00000\$05.00}
\authoraddr{Chun-yu Ma and Danny Summers,
Department of Mathematics and Statistics, 
Memorial University of Newfoundland, 
St John's, Newfoundland, A1C 5S7, Canada
(e-mail: cyma@math.mun.ca, dsummers@math.mun.ca)}

\slugcomment{Ref. {\it Geophysical Research
Letters 25}, No.21, 4099, 1998.}

\begin{document}

\title{Formation of Power-law Energy Spectra in Space Plasmas
       by Stochastic Acceleration due to Whistler-Mode Waves}

\author{Chun-yu Ma \altaffilmark{1}
        and Danny Summers }

\affil{Department of Mathematics and Statistics, Memorial University of 
Newfoundland, St John's, Newfoundland, A1C 5S7, Canada}

\altaffiltext{1}{
On leave from Purple Mountain Observatory, Chinese Academy of Sciences,
Nanjing, P.R. China.}

\begin{abstract} 
A non-relativistic Fokker-Planck equation for the electron distribution
function is formulated incorporating the effects of stochastic
acceleration by whistler-mode waves and Coulomb collisions. The stationary
solution $f$ to the equation, subject to a zero-flux boundary condition,
is found to be a generalized Lorentzian (or kappa) distribution, which
satisfies $f\propto v^{-2(\kappa+1)}$ for large velocity $v$, where
$\kappa$ is the spectral index. The parameter $\kappa$ depends strongly on
the relative wave intensity $R$. Taking into account the critical energy
required for resonance of electrons with whistlers, we calculate a range
of values of $R$ for each of a number of different space plasmas for which
kappa distributions can be expected to be formed. This study is one of the
first in the literature to provide a theoretical justification for the
formation of generalized Lorentzian (or kappa) particle distribution
functions in space plasmas. 
\end{abstract}

\begin{article}

\section{Introduction}

In the natural space environment, e.g., planetary magnetospheres, and the 
solar wind (and actually many other kinds of astrophysical objects),
plasmas are generally observed to possess a particle distribution function
with a non-Maxwellian high-energy tail. The
distribution function typically has a power-law tail in energy 
[distribution $\propto$ (particle energy)$^{-(\kappa+1)}$],
and frequently can be modeled by a generalized Lorentzian (kappa)
distribution. For a variety of observational data, see
{\it Vasyliunas} [1968], {\it Gosling et al.} [1981],
{\it Armstrong et al.} [1983], {\it Christon et al.} [1988],
{\it Divine and Garrett} [1983]
and references contained in {\it Summers and Thorne} [1991],
{\it Collier} [1993],
and {\it Mace and Hellberg} [1995]. 
Associated with the generalized Lorentzian distribution has been
developed the modified plasma dispersion function which is an effective 
tool for investigating waves and microinstabilities in space
plasmas \cite{ds1,mh,ds2,ds3}.
Physical mechanisms that can produce power-law particle distributions
include stochastic acceleration by plasma wave turbulence, and
collisionless shocks. 
Kappa-like distribution functions which are Maxwellian at low energies and
power law at high energies can also be produced by velocity-space L\'{e}vy
flight probability distributions [e.g., {\it Collier}, 1993]. 
It is the process of stochastic acceleration by plasma wave turbulence
with which we shall be concerned in this paper.
Many researchers have studied stochastic acceleration of charged
particles by various modes of plasma wave, e.g., 
{\it Gurevich} [1960], 
{\it Kennel and Engelmann} [1966], 
{\it Melrose} [1980, 1986], 
{\it Steinacker and Miller} [1992], and 
{\it Schlickeiser} [1997].
{\it Hasegawa et al.} [1985] showed that superthermal radiation
can enhance velocity-space diffusion so as to produce a
power-law distribution. However, these authors did not estimate values of
the intensity of the radiation field and other plasma parameters necessary
to obtain a reasonable value for the power law. The source of the free
energy to excite such intense radiation was another question left
unaddressed.
{\it Melrose} [1980] pointed out that the equilibrium
distribution is a power law only if the momentum diffusion coefficient
$D(v)$ is proportional to $v^{-1}$, where $v$ is the
particle velocity. For example,
for Langmuir waves, $D(v)\propto v^{-1}$; but this requires the
turbulent spectrum $I(k)$ to be independent of the wave number $k$,
which is an unlikely possibility [{\it Melrose}, 1980]. 
Stochastic acceleration by plasma waves is characteristically 
slow due to the diffusive nature of the scattering process.
Nevertheless, the process may be applicable, for instance, in planetary
magnetospheres in cases when typical electron acceleration times 
are of the order of tens of hours.  
We note that {\it Schlickeiser} [1997] has recently obtained a
diffusion coefficient varying as $D(v)\propto v^{-1}$ when the charged
particles are accelerated by whistler-mode waves. 
The momentum diffusion coefficient results from the
resonant interaction between particles and whistler-electron cyclotron
plasma waves. Similar formulae have also been obtained by {\it
Dermer et al.} [1996]. 
Numerical calculations of the momentum diffusion coefficients for
particle-whistler interaction were given by 
{\it Steinacker and Miller} [1992] and
{\it Pryadka and Petrosian} [1997],  
but no analytical formulae were obtained
for medium energetic electrons ($10$ keV to a few hundreds of keV),
which limits their application by other authors.
There are several physical processes that can drive the whistler
instability. For example, anisotropies in the plasma velocity
distribution can make whistlers unstable. Moreover,
a kappa distribution can itself enhance the whistler-mode
instability, as compared with the Maxwellian distribution 
[e.g., see {\it Summers and Thorne,} 1992; 
{\it Xue et al.,} 1993;
{\it Mace,} 1998].

In the following section, we shall develop a non-relativistic theory for
the formation of kappa distributions by means of whistler-mode wave
stochastic acceleration.

\section{Theory}
We assume a Kolmogorov-type magnetic turbulence power
spectrum in the
wave number for the whistler-mode waves, namely,
\begin{equation}
   I=I_0k_\parallel^{-q} , 
\end{equation}
with $q > 1$, where $k_\parallel \geq k_{min}$,
$k_\parallel$ is the parallel wave number, 
and $I_0=(q-1)(\delta B)^2k_{min}^{q-1}$ is the energy
density of the whistler turbulence. It is natural to assume that
$k_{min} V_A=\Omega_i$ for whistler waves, where $V_A$ and $\Omega_i$ are
respectively the Alfv\'en speed and the non-relativistic ion  
gyrofrequency. In this paper we are concerned with the situation in which
pitch-angle anisotropy is small and the particle distribution is
isotropic. For higher energy electrons, it is generally the case that
the time-scale for pitch angle diffusion is much less than that for
momentum diffusion, and so the assumption of isotropy is justified. For
lower energy electrons, where the pitch angle and momentum diffusion rates
are comparable, we assume that isotropization by Coulomb collisions
occurs. The momentum diffusion coefficient can then be averaged over the
pitch angle. Herein, we employ the diffusion coefficient given by
{\it Schlickeiser} [1997], namely,
\begin{equation}
  D=c^2\Omega_e{\pi(q-1)\over{8}}\left({\delta B\over
       B_0}\right)^2\beta_A^{2+q}
       \left({k_{min}c\over\Omega_i}\right)^{q-1}
       \left({m_p\over m_e}\right)J_W\beta^{-1} , \label{diff}
\end{equation}
where $\Omega_e$ is the non-relativistic electron
gyrofrequency, $\beta_A=V_A/c$, $\beta=v/c$, $c$ is the speed of light,
and $J_W$ is a weakly varying function of $v$.
Henceforth in this paper, the Kolmogorov turbulent spectrum ($q=5/3$)
is adopted. It then follows that $J_W$ is of order unity.
In accordance with the whistler
wave dispersion relation and the resonance condition \cite{mel1986}, 
we find that electrons can resonate with whistlers provided that 
$\beta \ge (m_p/m_e)^{1/2}\beta_A$.
The diffusion coefficient (\ref{diff}) is valid in the region
where $(m_p/m_e)^{1/2}\beta_A \le \beta \le (m_p/m_e)\beta_A$. 
However, lower-energy electrons can be
accelerated by high wave-number whistlers as the electron
cyclotron branches of the dispersion relation are approached\cite{p}.

Coulomb interactions result in both momentum diffusion and friction
in the kinetic equation for the particle distribution function.
As we are assuming the distribution to be isotropic, we can ignore
the angular contribution to Coulomb collisions \cite{hin},
giving the equation,
\begin{equation}
  {\partial f\over {\partial t}}={1\over v^2}
  {\partial\over{\partial v}}v^2\left(F_cf +
  {1\over 2}D_{\parallel c}{\partial f\over{\partial v}}\right),
 \label{coul} 
\end{equation}
where the frictional coefficient $F_c$ is given by
\begin{equation}
  F_c=\nu_s v+{D_{\parallel c} - D_{\perp c}\over v} +
       {1\over 2}{dD_{\parallel c}\over{dv}}.
\end{equation}
$\nu_s$ is the slowing-down rate, and $D_{\perp c}$ and
$D_{\parallel c}$ are the respective perpendicular and 
parallel diffusion coefficients.
When the velocity of the particle $v$ is much larger than the thermal
velocities of the electrons and ions 
[$v_e\ll v, v_i\ll v, v_{e,i}=(2T_{e,i}/m_{e,i})^{1/2}$], we have by
{\it Hinton} [1983],
\begin{equation}
  \nu_s=3n_e{\Gamma_e\over v^3}, 
\end{equation}
\begin{equation}
D_{\perp c}={2n_e\Gamma_e\over v} ,
\end{equation}
\begin{equation}
D_{\parallel c}={n_e\Gamma_e (v_e^2+v_i^2)\over v^3} 
               \simeq {n_e\Gamma_e v_e^2\over v^3} ,
\end{equation}
where $n_e$ is the electron number density,
\[
  \Gamma_e ={4\pi e^4 \log_{10} \Lambda\over{m_e^2}} ,
\]
and $\log_{10}\Lambda\simeq 20$ is the Coulomb logarithm.
Finally, modifying Equation (\ref{coul}) to include  acceleration by
whistler-mode waves, we obtain the following Fokker-Planck equation
to describe the evolution of the distribution function:
\begin{equation}
  {\partial f\over {\partial t}}={1\over v^2}
  {\partial\over{\partial v}}v^2\left(F_cf +
  {1\over 2}D_{\parallel c}{\partial f\over{\partial v}}\right) +
  {1\over v^2}{\partial\over{\partial v}}\left(v^2
  D{\partial f\over{\partial v}}\right) , \label{fv}
\end{equation}
where $D$ is given by Equation (\ref{diff}).
With a zero-flux boundary condition, the stationary solution of 
Equation (\ref{fv}) is given by
\begin{equation}
f = A\exp\left\{-\int{{2n_e\Gamma_e v^{-2}\over {n_e\Gamma_e
         v_e^2v^{-3}}+2D}}dv\right\}. \label{f1}
\end{equation}
Substituting the expression (2) for $D$ in Equation (\ref{f1}), we obtain
\begin{eqnarray}
f &=& A\exp\left\{ -\int{ { d[v^2/v_e^2]\over
      { 1+{ v^2\over{(\kappa+1) v_e^2} } } } } \right\} \cr
  & & \cr
  &=& A\left[1+{v^2\over {\kappa \theta^2}}
\right]^{-(\kappa+1)},\label{fs}
\end{eqnarray}
where 
\begin{equation}
  \kappa+1 = {24e^4n_e\log_{10}\Lambda \over
           {\Omega_e m_p m_e c^3 R \beta_A^3 J_W}}, \label{kp}
\end{equation}
$A$ is a constant of integration, $R=(\delta B/B_0)^2$ is the power of the
wave turbulence, and  
$\theta=[(\kappa+1)/\kappa]^{1/2}v_e$. Normalizing the distribution
using $\int_0^\infty{4\pi v^2 f dv} =1$ gives
\[
  A={1\over{(\pi\kappa)^{3/2}\theta^3}}
    {\Gamma(\kappa+1)\over{\Gamma(\kappa-1/2)}}
\]
where $\Gamma$ is the gamma function.
The distribution (\ref{fs}) is formally a generalized Lorentzian (kappa)
distribution, with the spectral index $\kappa$ given by Equation 
(\ref{kp}). 
Parenthetically, it can be noted that setting $D=0$ in Equation~(\ref{f1})
leads to the recovery of the Maxwellian distribution
$f=A\exp(-v^2/v_e^2)$.
From Equation (\ref{kp}) we determine that $\kappa$ is given by
\begin{equation}
  \kappa+1 = 0.9 \times 10^{-25}n_e^{5/2}B_0^{-4}R^{-1}J_W^{-1},
  \label{kvalue}
\end{equation}
where $n_e$ is in cm$^{-3}$, and $B_0$ in Gauss.

\section{Discussion}
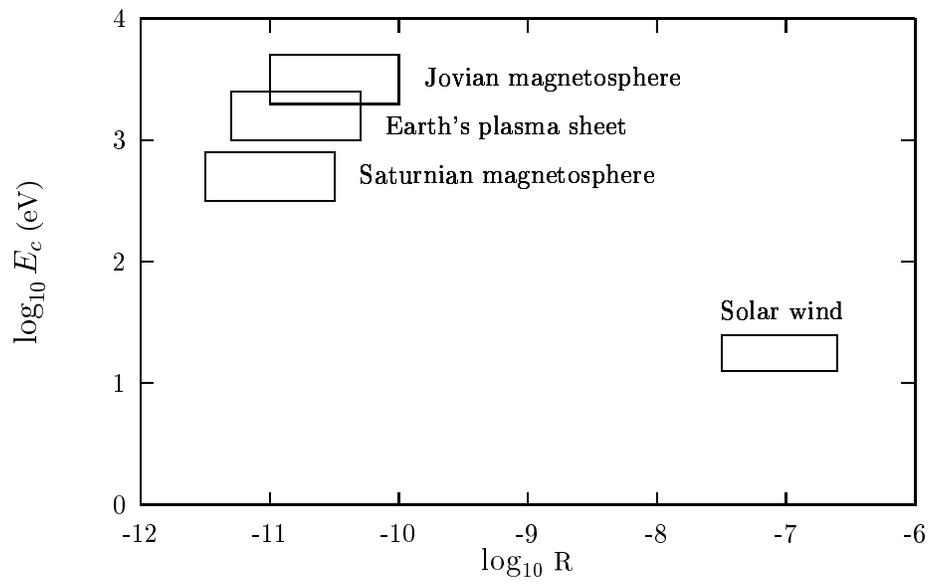
\begin{figure}
\caption{The range of values of the relative wave intensity $R$ ,
calculated from Equation (\ref{kvalue}), corresponding to the range of
$\kappa$-values, $2\le\kappa\le 8$, for space plasmas for which typical
parameters and $E_c$-values are given in Table 1.
} 
\label{fig}
\end{figure}

In the foregoing,
we have determined that stochastic acceleration
by whistler-mode turbulence can produce a particle distribution with a
power-law tail, i.e., $f\propto v^{-2(\kappa+1)}$. 
By Equation~(\ref{kvalue}), we note that
the value of the spectral index $\kappa$ depends strongly on the relative
wave intensity $R$.
In 
\callout{Table~\ref{tbl-1}},
for various space plasmas, we have listed typical
values of the parameters $n_e, B_0, \beta_A$ and the thermal
energy(temperature). 
The basic plasma parameters in this table for the  
solar wind were taken from {\it Kivelson and Russell} [1995], 
for the Earth's plasma sheet from {\it Christon et al.} [1988],
for the Jovian magnetosphere from {\it Dessler} [1983] and
{\it Divine and Garrett} [1983], and
for the Saturnian magnetosphere from {\it Sittler et al.} [1983].
Because only the energetic electrons with velocities
satisfying $\beta \ge (m_p/m_e)^{1/2}\beta_A$ can resonate with
first-harmonic whistlers, we list in Table 1
the critical energy $E_c$ corresponding to
the critical velocity $\beta_c=(m_p/m_e)^{1/2}\beta_A$.
If the critical energy $E_c$ is much larger than the electron thermal
energy, then the whistlers can only accelerate non-thermal high energy
particles, for which the Coulomb momentum diffusion coefficient 
$D_{\parallel c}$ is much less than the acceleration diffusion
coefficient. In this case, an injection mechanism is required to
pre-accelerate the particle to the required critical energy $E_c$. 
It is straightforward to show that this situation applies, for instance, 
to solar flares.
In fact, there is no observational evidence that electrons have a
kappa distribution in solar flares.
If the critical energy $E_c$ is of the same
order as the thermal energy, then whistlers can accelerate the thermal
particles into the high-energy tail of the distribution, and so
a kappa distribution can be produced. This is essentially the
case for all the space plasmas shown in 
\callout{Figure~\ref{fig}}.
For a specific plasma characterized by its electron number density $n_e$
and background magnetic field $B_0$, formula (\ref{kvalue}) gives the
spectral index $\kappa$ produced by whistler turbulence of power $R$. In
Figure ~\ref{fig}, for each of the given space plasmas, we show the range
of $R$-values associated with the range of $\kappa$-values that might be
typically observed, namely $2\le\kappa\le 8$. In addition, we also show
the value of the critical energy $E_c$ for each plasma. 
The results imply that relatively weak whistler-mode turbulence could
produce power-law spectra in the Earth's plasma sheet and in the Jovian
and Saturnian magnetospheres,  whereas stronger turbulence is required in
the solar wind. 

In conclusion, we should point out both the merits and the limitations of
this study. Of course, it is not the case that all electron power-law
spectra are produced by whistler-mode turbulence, or that whistler-mode
turbulence will always give rise to electron power-law spectra.
Nevertheless, under certain restrictions, we have shown that the
production of electron power-law energy spectra by stochastic acceleration
due to whistlers is a viable possibility. Apart from the non-relativistic
fromework, limitations of the study include the assumptions of isotropy, a
zero-flux boundary condition, and a momentum diffusion coefficient based
on wave propagation strictly parallel to the background magnetic field.
The intrinsic value of the study is that it is one of the first in the
literature to present a firm theoretical basis for the formation of
generalized Lorentzian (or kappa) particle distribution functions in space
plasmas.

\acknowledgments
This work is supported by the Natural Sciences and Engineering Research
Council of Canada under Grant A-0621. Additional support is acknowledged
from the Dean of Science, Memorial University of Newfoundland.
The paper was completed when
D. S. was Visiting Professor at the Radio Atmospheric Science Center,
Kyoto University, Japan. It is a pleasure to thank Professor Hiroshi
Matsumoto of Kyoto University for his generous hospitality.  

{}

\end{article}
 
\clearpage

\clearpage

\begin{planotable}{lcccc}
\tablewidth{41pc}
\tablecaption{Typical parameters for different space plasmas. Also given
are values for the critical energy $E_c$ which is the minimum
particle energy required for particle acceleration by whistler-mode
waves.}
\tablenum{1}
\tablehead{
      \colhead {Plasma parameters} & 
      \colhead {Solar wind\tablenotemark{a}} &
      \colhead {Earth's plasma sheet\tablenotemark{b}} &
      \colhead {Jupiter \tablenotemark{c}} &
      \colhead {Saturn\tablenotemark{d}}
     }
\tablenotetext{a}{At 1 AU, {\it A.J. Hundhausen,}
   in {\it Kivelson and Russell} [1995], page 92.}
\tablenotetext{b}{At L=14, {\it Christon et al.} [1988].}
\tablenotetext{c}{At L=6, e.g., {\it Dessler} [1983] and 
                  {\it Divine and Garrett} [1983].}
\tablenotetext{d}{At L=15, {\it Sittler et al.} [1983].}

\startdata
\label{tbl-1}
Density $n_e$ (cm$^{-3}$)        
  &     7       & 1   & $2~10^3$  & 0.14 \nl
Magnetic field $B_0$ (G)       
  & $7~10^{-5}$ & $ 2~10^{-4}$ & $1.8~10^{-2}$ & $6~10^{-5}$ \nl
Alfv\'en speed parameter $\beta_A$        
  & $1.9~10^{-4}$ & $1.5~10^{-3}$ & $2.8~10^{-3}$  & $1.1~10^{-3}$ \nl
Thermal energy (eV)
  & 12 & 1000 & 1000 & 300 \nl
\\[-1.6ex]
\tableline
Critical energy $E_c$ (eV)
  & 17 & 1500  & 3500 & 500 
\end{planotable}

\clearpage

\end{document}